\documentclass[12pt]{iopart}

\usepackage{iopams} 
\usepackage{here}

\usepackage{graphicx}
\begin{document}

\title[Short-range test of the universality of gravitational constant $G$ at the millimeter scale using a digital image sensor]{Short-range test of the universality of gravitational constant $G$ at the millimeter scale using a digital image sensor}

\author{K. Ninomiya, T. Akiyama, M. Hata, M. Hatori, T. Iguri, Y. Ikeda, S. Inaba, H. Kawamura,R. Kishi, H. Murakami, Y. Nakaya, H. Nishio, N. Ogawa, J. Onishi, S. Saiba, T. Sakuta, S. Tanaka, R. Tanuma, Y. Totsuka, R. Tsutsui, K. Watanabe and J. Murata}

\address{
Department of Physics, Rikkyo University, 3-34-1 Nishi-Ikebukuro, Tokyo 171-8501, Japan}
\ead{kazufumi@rikkyo.ac.jp}
\begin{abstract}
 The composition dependence of gravitational constant $G$ is measured at the millimeter scale to test the weak equivalence principle, which may be violated at short range through  new Yukawa interactions such as the dilaton exchange force. A torsion balance on a turning table with two identical tungsten targets surrounded by two different attractor materials (copper and aluminum) is used to measure gravitational torque by means of digital measurements of a position sensor. Values of the ratios
$\tilde{G}_{Al-W}/\tilde{G}_{Cu-W} -1$ and $\tilde{G}_{Cu-W}/G_{N} -1$ were $(0.9 \pm 1.1_{\mathrm{sta}} \pm 4.8_{\mathrm{sys}}) \times 10^{-2}$ and $ (0.2 \pm 0.9_{\mathrm{sta}} \pm 2.1_{\mathrm{sys}}) \times 10^{-2}$ , respectively; these were obtained at a center to center separation of 1.7 cm and surface to surface separation of 4.5 mm between target and attractor, which is consistent with the universality of $G$.
A weak equivalence principle (WEP) violation parameter of
$\eta_{Al-Cu}(r\sim 1\: \mathrm{cm})=(0.9 \pm 1.1_{\mathrm{sta}} \pm 4.9_{\mathrm{sys}}) \times 10^{-2} $ at the shortest range of around 1 cm was also obtained.
\end{abstract}

\submitto{\CQG}
\section{INTRODUCTION \label{intro}}
The universality of gravitational constant $G$ (UGC) and that of free fall (UFF) are consequences of the weak equivalence principle (WEP), which states that the ratio between gravitational mass $m_{g}$ and inertial mass $m_{I}$ is independent of material composition
\cite{fischbach1999search,will1993theory,damour1996testing}.
Although WEP is a fundamental principle in gravitational physics, several theoretical models have predicted its violation through, for instance, the dilaton exchange force 
\cite{damour2012theoretical,dent2008eotvos}.
On the contrary, recent experiments such as the E\"ot-Wash experiment 
\cite{schlamminger2008test,0264-9381-29-18-184002} and lunar laser ranging measurements 
\cite{williams2004progress}
report a $10^{-13}$ level confirmation of the composition independence of gravitational acceleration.
Although WEP has been sufficiently tested at the large scale over km range, its validity for the short range, where a possible new boson exchange force can be probed as an additional interaction, should be tested  
\cite{schlamminger2008test}.
This study aims to examine a new short range interaction by testing UGC at the millimeter scale, a region wherein no experimental test of WEP has yet been conducted.

A modified gravitational force between objects $i$ and $j$, with an additional new term can be expressed as
\begin{equation}
F(r)=G_N\frac{m_{i}m_{j}}{r^2}+G_N \frac{\tilde{m}_{i}\tilde{m}_{j}}{r^2} a(r)=\tilde{G}_{ij}(r)\frac{m_{i}m_{j}}{r^2},
\label{GENERAL}
\end{equation}
where $G_{N}$ is the Newtonian gravitational constant and $a(r)$ is distance dependence factor of the additional force term.
The additional term is proportional to a new ``mass-like" point charge $\tilde{m}$, which is analogous to the usual gravitational mass $m_g$.
In previous WEP tests, a generalized point charge $q=\tilde{m}/u$ is regarded as a function of neutron number and proton number, where $u$ is the atomic mass, has often been used. As WEP is well tested at a high precision of $10^{-13}$ at a planetary scale 
\cite{roll1964equivalence,braginsky1972equivalence,will2014confrontation} (i.e., for Earth), we can assume $a(r) \rightarrow 0 $ $(r \rightarrow \infty)$ in this study.
If we observe in a short range experiment that $\tilde{m}$ is not equal to $m_g$ (which is measured at long distance),  then UFF must be violated at short range.
This can be assumed because the known ``mass" is determined using Earth's gravity, at a scale wherein WEP is well tested, and therefore they can be regarded as the inertial mass within the precision of long range WEP tests.

Modification of the gravitational force can be tested in terms of the modified gravitational constant
\begin{equation}
\tilde{G}_{ij}(r)=G_N \left( 1+\frac{q_i}{m_i/u} \frac{q_j}{m_j/u} a(r)\right)
=G_N \left( 1+A_{ij} a(r)\right).
\label{mod-G}
\end{equation}
Here, a composition dependence modification factor
\begin{equation}
A_{ij}=\frac{q_i}{m_i/u} \frac{q_j}{m_j/u}
\end{equation}
is introduced for simplicity.
Under the condition of WEP, $A_{ij}=1$ for all of $i,j$.

In this study, we aim to investigate the possibilities that $A_{ij}\neq 1$ and $a(r)\neq 0$.
To this purpose, the value of $\tilde{G}$ at different combinations of materials was measured at the millimeter scale, which cannot be done using short-range inverse square law tests such as those in 
\cite{PhysRevLett.98.021101,PhysRevLett.98.131104}, without directly testing the composition dependences of different materials.
Both $A_{ij} \neq 1$ (violation of the universality of free fall) and $a(r)\neq 0$ (violation of the inverse square law) are required to deduce a composition dependence of $\tilde{G}$.

As discussed in our recent review \cite{CQG-Review},
when testing an inverse square law without consideration of composition dependence, the Yukawa force is widely used to represent $a(r)$ with its short interaction range $\lambda$ of the new interaction and coupling strength $\alpha$ as
\begin{equation}
a(r)=\alpha \left(  1+\frac{r}{\lambda}\right) e^{-r/\lambda}.
\label{ar-Yukawa}
\end{equation}
However, other models, such as the large extra-dimension model  \cite{Arkani.Hamed1998263}, obey a modified power law force instead of the single Yukawa force. In such cases, a power law force with a characteristic distance $\lambda$ and new power parameter $n$,
\begin{equation}
a(r)=(1+n)\left( \frac{\lambda}{r} \right)^n
\label{ar-power}
\end{equation}
is preferable to be used for describing the wide dynamic range of $\lambda$, especially $\lambda$ at distances significantly greater than the experimental test distance $r$ \cite{CQG-Review}.

In this study, we propose to extend these parametrizations to composition-dependent analysis.
Details regarding the interpretation of the experimental results of $\tilde{G}$ in the model parameter spaces will be discussed in Section IV.

\begin{figure}
\begin{center}
\includegraphics[clip,width=100mm]{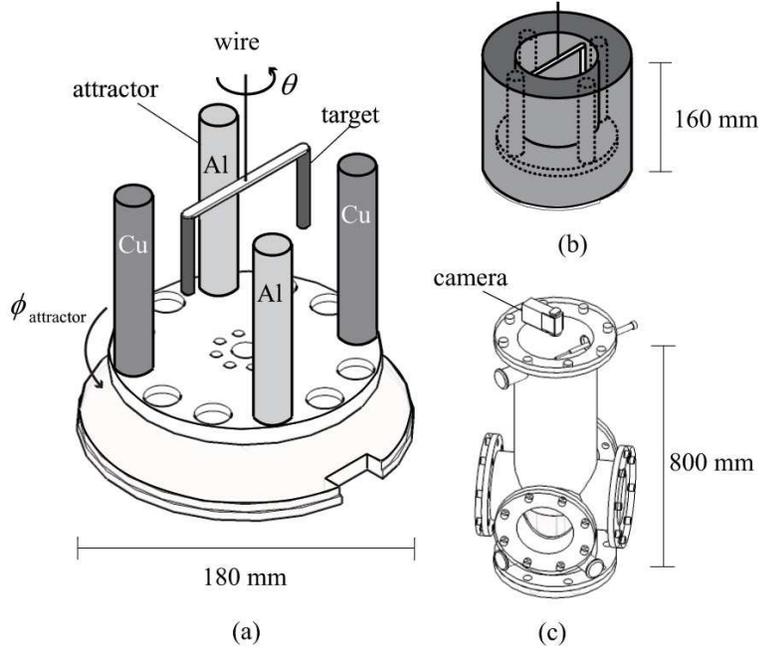}%
\caption{\label{apparatus} Experimental apparatus (Newton-II) comprising a torsion balance hung from a wire and attractors (a), which are surrounded by an electric shield cover (b). The entire setup is placed inside a vacuum chamber (c). Two tungsten targets are attached to the ends of the torsion balance. Two aluminum and two copper attractors are positioned on a turning table.}\end{center}
\end{figure}

\section{EXPERIMENT \label{exp}}
Figure \ref{apparatus} shows the experimental apparatus ``Newton-II," designed to measure gravitational torque from attractors of different compositions on a torsion balance.
The torque signal is obtained as the twisting angle of the torsion balance $\theta$, which is visually monitored by a position sensor using an online digital-image-analysis system
\cite{muratapico,hata2009recent,ninomiya2009new,ninomiya2013short}.

During a measurement, the angular position of the attractor $\phi_{attractor}$ slowly rotates around the torsion balance, while monitoring $\theta$. 
As the time scale of gravity changing due to the attractor rotation is considerably larger than the free torsional oscillation period of the torsion balance, the balanced angular position between gravity and the torsional spring force can be measured as a synchronized signal with the attractor rotation.

The torsion balance comprises two tungsten columns (targets) suspended on both ends of an aluminum bar, 
which is hung from a 30 $\mathrm{\mu m}$ diameter, 45-cm-long gold plated tungsten wire. 
We assume Hooke's law $\tau = -\kappa\Delta\theta$, where $\tau$ is the torque and $\kappa$ is the torsional spring constant, governs the wire twisting behavior.
Two copper and two aluminum attractor columns are placed parallel to the targets on a turning table, whose axis of rotation is the same as the target center axis. 
The details of the torsion balance and attractor components  are shown in Table \ref{detail}.
\begin{table} 
\begin{center}
\begin{tabular}{l}
 \hline
 \hline
 wire (gold plated tungsten)\hspace{12mm}$Dia.:30\: \mathrm{\mu m}$ \\
 			  \hspace{53mm} $L:450.0\pm 0.5\: \mathrm{mm}$ \\
 \hline
 target (tungsten)   \hspace{26mm}$Dia.:5.98\pm 0.03\: \mathrm{mm}$ \\
                     \hspace{54mm}$L:50.00\pm 0.02\: \mathrm{mm}$ \\
  \hspace{8mm} $two\; targets\; center\; to\; center\; dist.$ $:88.12\pm 0.04\: \mathrm{mm}$ \\ 	   
 \hline          
 torsion balance bar (aluminum)   \hspace{8mm}$W:2.11\pm 0.02\: \mathrm{mm}$ \\
			\hspace{54mm}$L:94.10\pm 0.02\: \mathrm{mm}$ \\
			\hspace{54mm}$T:6.02\pm 0.02\: \mathrm{mm}$ \\
 \hline                        
 attractor (copper or aluminum)	   \hspace{6mm}$Dia.:20.00\pm 0.02\: \mathrm{mm}$ \\
 			  	   \hspace{54mm}$L:116.40\pm 0.03\: \mathrm{mm}$  \\
 \hspace{25mm} $pitch\; circle\; diameter$ $:123.00\pm 0.04\: \mathrm{mm}$\\
 \hline
 \hline
 \end{tabular}
 \end{center}
 \caption{\label{detail} Details of the torsion balance and attractor components.}
 \end{table}
To eliminate the influence of electric fields on the target, the attractor is surrounded by an electrical shield cover made of copper. 
All apparatus components are electrically conductive and made of non-magnetic metals, and the unit is mounted inside a vacuum chamber.
The vacuum level is maintained at around 1 Pa; the vacuum pumps do not operate during the measurements to avoid the influence of mechanical vibrations.
The attractor turning table is rotated using a stepping motor with a rotational speed of 360 degrees per 5 hours, which is digitized in 0.005 degree steps. 
The angle of rotation $\theta$ is measured using a CCD camera, positioned outside the vacuum chamber, which views the assembly through an acrylic viewport at the top of the chamber. 
The shortest distances between the target and each attractor are 1.7 cm center to center and 0.4 cm surface to surface.  
To avoid mechanical noise, the apparatus is set in a basement room at Rikkyo University. 
The attractors move near the outer region of the targets, enabling us to maintain rotation of the attractors around the torsion balance.

\begin{figure}
\begin{center}
\includegraphics[clip,width=100mm]{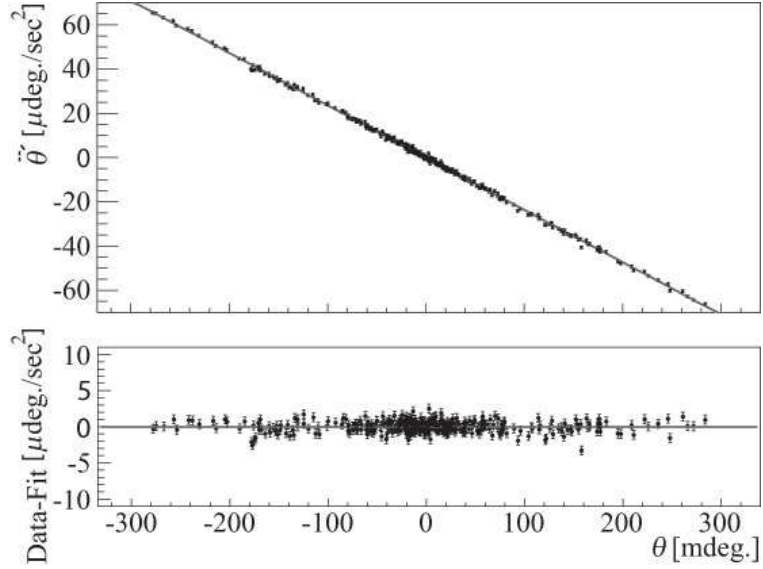}
\caption{\label{Hooke} Corrected angular acceleration $\ddot{\theta}'=\ddot{\theta}+\gamma/I_{target}\cdot\dot{\theta}$ is plotted as a function of $\theta$ (top). The solid line indicates the linear correlation function $\ddot{\theta}'=-\kappa/I_{target}\cdot\theta$ expected from Hooke's law.
The residual between them is also shown (bottom).}
\end{center}
\end{figure}

Our reliance on Hooke's law is examined by testing the deviation from harmonic oscillation in a free oscillation measurement without moving the attractors. 
The resulting free oscillation data were compared with the outputs of the torsional equation of motion $I_{target}\ddot{\theta}+\gamma\dot{\theta}+\kappa\theta=\tau_{external}$, where $I_{target}$ is the inertial moment of the target and $\gamma$ is the coefficient of friction.
Figure \ref{Hooke} shows the correlation, which should be linear and negative under Hooke's law, between $\theta$ and its acceleration obtained by second order time differentiation of $\theta$.
The influence of the friction term is eliminated in Figure \ref{Hooke}, in which the corrected angular acceleration $\ddot{\theta}'=\ddot{\theta}+\gamma/I_{target}\dot{\theta}$ is plotted, showing a clear linear correlation at $|\theta|\lesssim$ 0.3 degrees.
From this correlation and using a calculated inertial moment of $I_{target}=(1.08\pm 0.01)\times 10^{-4}\: \mathrm{Nms^2/rad}$, a torsional spring constant of $\kappa=(2.61\pm 0.03)\times 10^{-8}\: \mathrm{Nm/rad}$ is obtained.
Thus, the systematic error is estimated to be less than 1\% in $\kappa/I_{target}$. 
The torsional oscillation period is $T=403.55\,  \pm\,  0.02$ sec and amplitude damping life time is $6913 \pm 31$ sec.

The video data capture system comprises a CCD camera and PCI video capture board. 
Instead of performing offline extraction analysis of position-information data retrieved from image data recorded on a disk, the image data are buffered on a capture board memory that is accessed during the data-collection process; thus, information pertaining to only the torsion balance position is calculated and recorded.
Very high positional resolution better than the optical resolution or pixel size limit is obtained, as the position determination precision corresponds not to the standard deviation but to the standard error of the center of gravity of the position distribution \cite{muratapico}. 
$\theta$ is determined by performing a linear line fitting for the center-of-gravity position sequence for every video frame independent of the parallel pendulum motion of the torsion balance.
The angle $\theta$ is measured as a function of the continuously rotating $\phi_{attractor}$. 
This configuration is designed to suppress systematic error and maximize sensitivity to the relative strength of gravitational force for different materials. 
For example, the zero positions of $\theta$ and $\phi_{attractor}$ can be determined from the data obtained using the symmetrical configuration, without performing dedicated additional measurements.

\begin{figure}
\begin{center}
\includegraphics[clip,width=100mm]{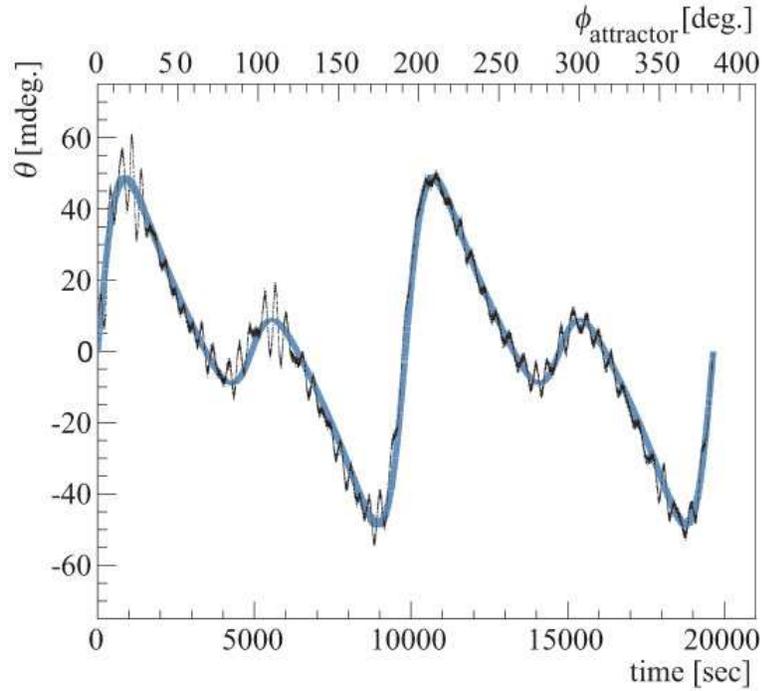}
\caption{\label{Tdata} A typical time sequence of $\theta$ for two cycles. Statistical errors are shown in black data points; Newtonian prediction with systematic errors is shown as the shaded band. Free torsional oscillation can be seen as the fast oscillations of the black dots around the shaded area, which is removed by a high-frequency filter in later analysis. Time-drifting effects have already been corrected in this plot.}
\end{center}
\end{figure}

\section{RESULTS \label{results}}
\begin{figure}
\begin{center}
\includegraphics[clip,width=100mm]{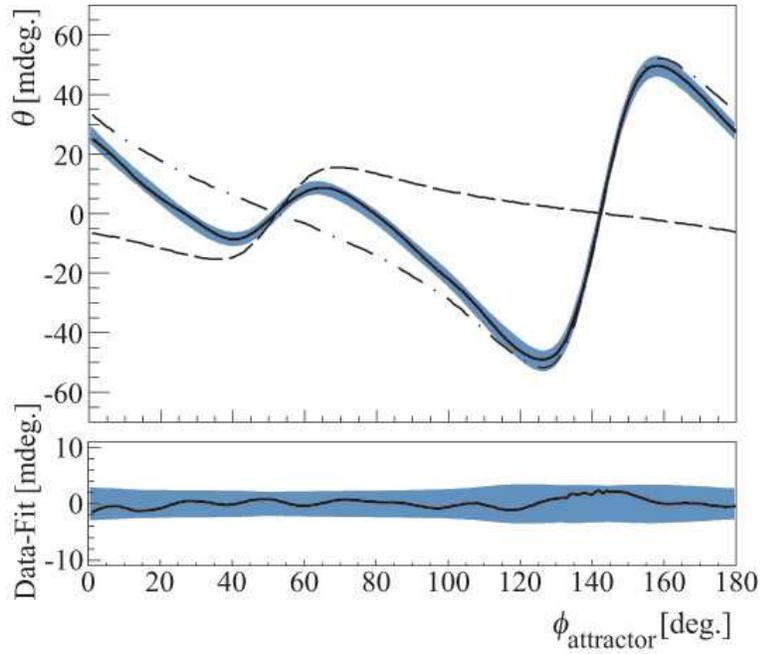}
\caption{\label{result} Superposition of $\theta$ from all  accumulated data is plotted as a function of $\phi_{attractor}$ (top). The broken and dot-dashed lines show the contributions from aluminum and copper attractors, respectively. Statistical errors are shown in the black data points, and the Newtonian prediction with systematic errors is shown as the shaded band. The residual between them is also shown (bottom).}
\end{center}
\end{figure}

A typical time sequence result is shown in Figure \ref{Tdata}, wherein $\theta$ is plotted as function of time, which is proportional to $\phi_{attractor}$. 
Figure \ref{Tdata} clearly shows a superposition of large and small oscillations, corresponding to the gravitational torque, mainly from the copper or aluminum attractors.
In total, 140 hours of data are accumulated and superimposed after high-frequency filtering and time-drifting correction. 
The result of the superposition is shown in Figure \ref{result}. 

Systematic errors $\sigma_\theta^{sys}$ on $\theta$ resulting from electric, magnetic, and thermal influences are estimated by dedicated measurements. 
An artificial strong electric field, magnetic field, and temperature variation are applied while monitoring the twisting effects without moving the attractors; this is compared with the real environment to estimate their remaining effects after experimentally minimizing them.
The obtained systematic error budget is shown in Table  \ref{error} along with the statistical resolution of the position sensor including thermal noises.
Note that the precision of this measurement is dominated by  temperature variation.
In Figures \ref{Tdata} and \ref{result}, the statistical error and all systematic errors, including the reliability of Hooke's law, are shown.
To enable a comparison with the experimental data after high-frequency filtering, the same filtering process is applied to the numerical calculation results.
The obtained results are consistent with the Newtonian calculation within the experimental errors.


 \begin{table} 
 \begin{center}
 \begin{tabular}{lll}
 \hline
 \hline
 systematic error & value & $\sigma_\theta^{sys}$ \\
 \hline
 magnetic effect  & $< 0.15\: \mathrm{\mu T}$	& $< 6.0 \times 10^{-4}\:  \mathrm{deg.}$ \\
 electric effect & $< 1\: \mathrm{mV}$	& $< 2.0 \times 10^{-9}\:  \mathrm{deg.}$ \\
 thermal effect	 & $< 0.58\: \mathrm{^oC}$	& $< 2.0 \times 10^{-3}\:  \mathrm{deg.}$ \\
 \hline
 mass ambiguity  	& 	&  \\
 target 	& $<0.78\: \mathrm{g}$	& $<9.9 \times 10^{-5} \: \mathrm{deg.}$ \\
 attractor 	& $<0.71\: \mathrm{g}$	& $<1.7  \times 10^{-5} \: \mathrm{deg.}$ \\
 \hline
 tilting ambiguity	&	\\
 target 	& $<0.25\: \mathrm{deg.}$ & $<1.7\  \times 10^{-4}\:  \mathrm{deg.}$\\
 attractor 	& $<0.01\: \mathrm{deg.}$ & $<4.0 \times 10^{-5} \: \mathrm{deg.}$\\
 \hline 
 misalignment &	\\
 vertical	& $<2.0\: \mathrm{mm}$ & $<8.4  \times 10^{-5}\: \mathrm{deg.}$\\
 horizontal	& $<0.5\: \mathrm{mm}$ & $<3.4  \times 10^{-4}\: \mathrm{deg.}$\\	
 \hline
 \hline
statistical precision && $\sigma_\theta^{sta}\sim2.6 \times 10^{-5} \: \mathrm{deg.}$ \\
 \end{tabular}
 \end{center}
 \caption{\label{error} Experimental error budget for systematic errors $\sigma_\theta^{sys}$ and statistical error $\sigma_\theta^{sta}$ are listed as typical values estimated at $\phi_{attractor}\sim$ 60 degrees. Systematic errors are included as parameter errors in the numerical calculation.} 
 \end{table}

\begin{figure}
\begin{center}
\includegraphics[width=100mm]{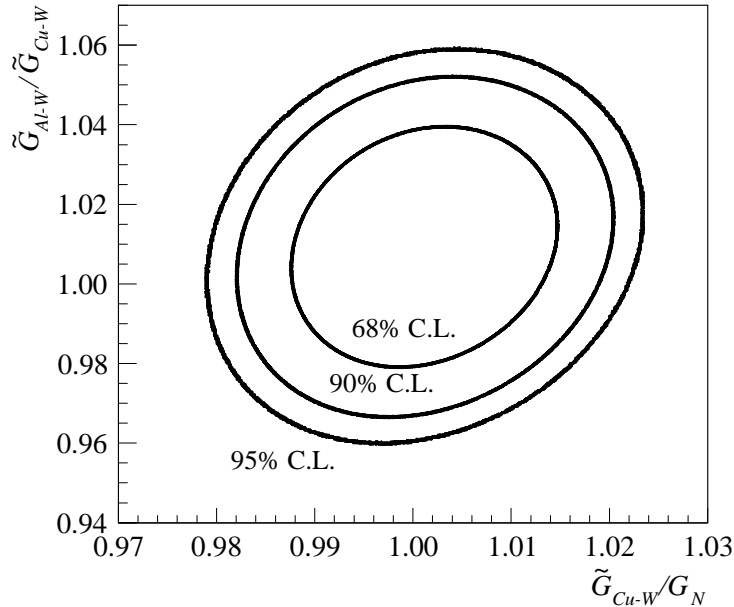}
\caption{\label{CDG} Composition dependence of the gravitational constant $G$ at $r \sim 1 \mathrm{cm}$. Optimized region of ratios between $\tilde{G}_{Al-W}$ (aluminum and tungsten), $\tilde{G}_{Cu-W}$ (copper and tungsten) and $G_N$ are plotted for 68 \%, 90 \%, and 95 \% confidence levels \cite{ninomiya2013short}.}
\end{center}
\end{figure}

The results are compared with the numerical calculation results with two compositions depending on the gravitational constants $\tilde{G}_{Al-W}$ (between aluminum and tungsten) and $\tilde{G}_{Cu-W}$ (between copper and tungsten) as free parameters, which are assumed to be constants over the present experimental length range. The optimized values are then obtained using a least square analysis, the result of which is shown in Figure \ref{CDG} using two ratios $\tilde{G}_{Al-W}/\tilde{G}_{Cu-W}$ and $\tilde{G}_{Cu-W}/G_{N}$. 
Here, the PDG (Particle Data Group) value of $G_{N}$ \cite{PDG} is used, and the ratios at $95\%$ confidence levels are obtained at $r \sim 1 \mathrm{cm}$ as follows:
\begin{eqnarray*}
\tilde{G}_{Al-W}/\tilde{G}_{Cu-W} -1 &=& (0.9 \pm 1.1_{\mathrm{sta}} \pm 4.8_{\mathrm{sys}}) \times 10^{-2}\\
\tilde{G}_{Cu-W}/G_{N} -1 &=& (0.2 \pm 0.9_{\mathrm{sta}} \pm 2.1_{\mathrm{sys}}) \times 10^{-2},
\end{eqnarray*}
which are consistent with UGC within the experimental precision. 
In addition, the obtained results show that the absolute values are consistent with known $G_{N}$, as
\begin{eqnarray*}
\tilde{G}_{Al-W} &=& (6.73 \pm 0.07_{\mathrm{sta}} \pm 0.32_{\mathrm{sys}}) \times 10^{-11} \; {\rm m^3/kg/s^2}\\
\tilde{G}_{Cu-W} &=& (6.69 \pm 0.06_{\mathrm{sta}} \pm 0.14_{\mathrm{sys}}) \times 10^{-11} \; {\rm m^3/kg/s^2}.
\end{eqnarray*}
This study confirms UGC at the shortest range of of around 1 cm for the first time in a direct measurement.


The obtained result can be interpreted as a WEP test by assuming that inertial mass $m_I$ is equal to gravitational mass $m_g$ measured at a long distance, where WEP is well confirmed as
\begin{equation}
\eta_{ij}=2\frac{(m_g/m_I)_i-(m_g/m_I)_j}{(m_g/m_I)_i+(m_g/m_I)_j}
<10^{-12}
\label{eta}
\end{equation}
 at $r>10^7$m  \cite{schlamminger2008test,williams2004progress}, using the WEP violation parameter $\eta$. Indeed, it can be shown that 

\begin{equation}
\eta_{ij}(r)=2\frac{\tilde{G}_{ik}(r)/\tilde{G}_{jk}(r)-1}{\tilde{G}_{ik}(r)/\tilde{G}_{jk}(r)+1}
\label{eta-G}
\end{equation}
if $m_g\rightarrow m_I$ at $r\rightarrow\infty$ for compositions $i$, $j$, and $k$. Our results can be expressed as follows:
\begin{equation}
\eta_{Al-Cu}(r\sim 1\: \mathrm{cm})=(0.9 \pm 1.1_{\mathrm{sta}} \pm 4.9_{\mathrm{sys}}) \times 10^{-2}.
\label{eta-exp}
\end{equation}
The present constraint on the WEP violation parameter is obtained at the shortest test scale of around 1 cm.

\section{DISCUSSION}
\label{discussion}

\subsection{model independent analysis}\label{discussion1}

\begin{figure}
\begin{center}
\includegraphics[clip,width=100mm]{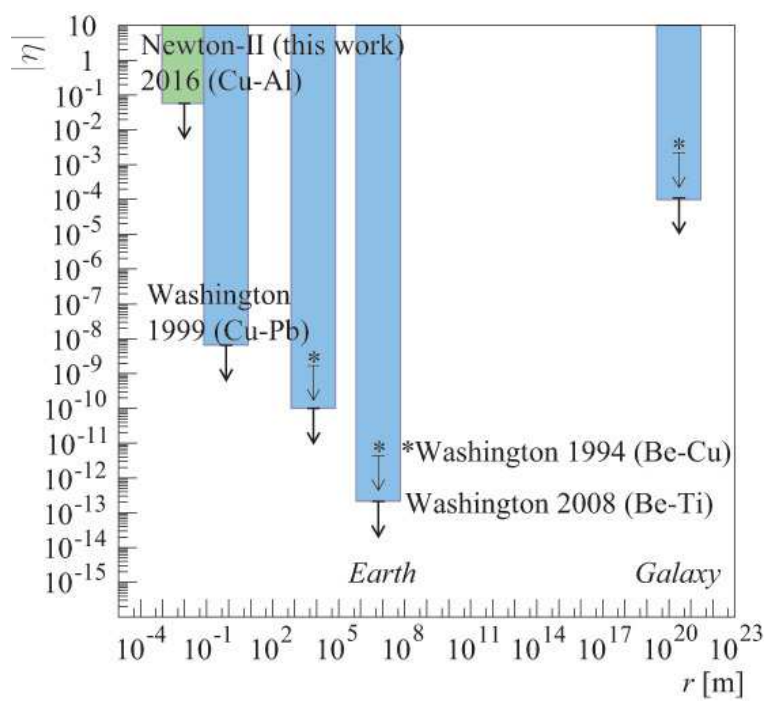}
\caption{\label{eta-r} Constraints on the WEP violation parameter $\eta$ plotted as a function of the measuring distance between two attracting objects. The result of this study is shown as Newton-II 2016, with Washington 1999 \cite{Smith.PhysRevD.61.022001}, 
1994 \cite{Su.PhysRevD.50.3614} and 2008 \cite{schlamminger2008test} results.}
\end{center}
\end{figure}

The obtained results on the WEP violation parameter $\eta$ is compared with results from other experiments, as shown in Figure \ref{eta-r}. As $\eta$ is defined as an experimental asymmetry of the gravitational constant between two objects with different compositions, this quantity does not require any model parameterization of the modified gravitational potential. In this sense, this  $\eta$ analysis is model-independent. As shown in Figure \ref{eta-r}, a very strong constraint on the upper limit on $\eta$ on the order $10^{-13}$ is obtained at a length scale of $r \sim 1000$ km. On the contrary, the previous constraints are very weak, both at a very long scale (proportional to the radius of the Milky Way galaxy) and at a short-range scale. Among these results, the present result sets a new constraint at the shortest range, although with low precision.

\begin{figure}
\begin{center}
\includegraphics[clip,width=100mm]{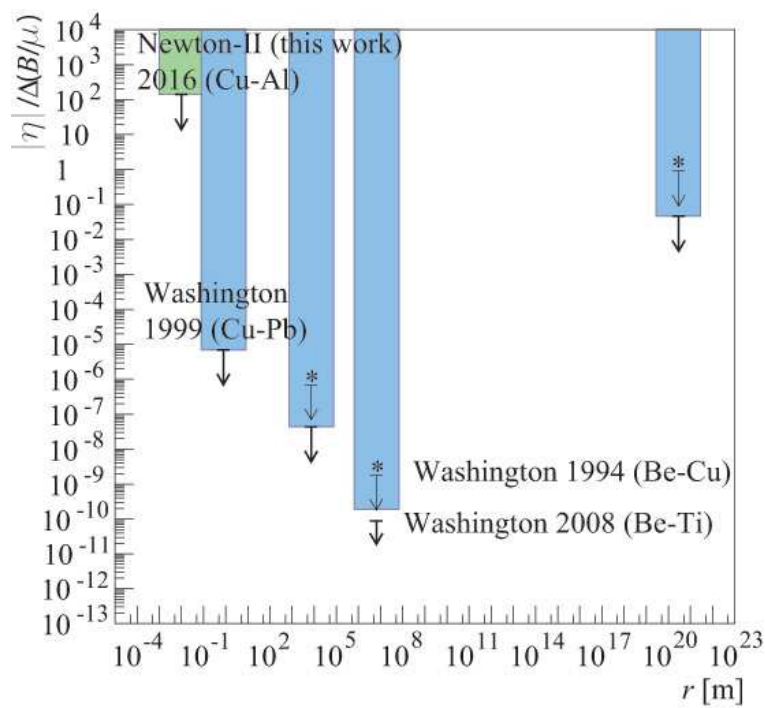}
\caption{\label{reduced-eta-r} Constraints on the ``reduced" WEP violation parameter $\eta/\Delta(B/\mu)$, plotted as a function of measuring distance. The result of this study is shown as Newton-II 2016. References are same as Fig.\ref{eta-r}.}
\end{center}
\end{figure}

The results in Figure \ref{eta-r} were obtained for various combinations of materials $i,j$. As such, it is not easy to directly compare the implications for different matter combinations; thus, the authors propose a new quantity, the ``reduced WEP violation parameter" $\eta'$, which is defined as
\begin{equation}
\eta'=\frac{\eta_{ij}}{\Delta(B/\mu)_{ij}},
\end{equation}
for various materials $i$ and $j$, where $B=Z+N$ is baryon number, $\mu=m/u$ is mass in atomic mass unit $u$. Using this ``normalization", the constraints on $\eta_{ij}$ can be compared for experiments performed with different materials. It is because it can be shown that $\eta_{ij}\sim \Delta(B/\mu)_{ij} a(r)$, therefore, $a(r)$ can be extracted by this definition. The results are shown in Figure \ref{reduced-eta-r}. As with the results for $\eta$, the present study sets a new constraint at the shortest scale, although the relative upper limit of $\eta/\Delta(B/\mu)$ increases mainly because of the small value of $\Delta(B/\mu)$ for aluminum and copper used in this experiment.

Figure \ref{reduced-eta-r} represents the normalized experimental constraints on the WEP violation, which are represented as measuring distances. Any theoretical model proposing WEP violation must be consistent with these data.

\subsection{model dependent analysis}\label{discussion2}

The results can also be interpreted in the parameterization of the conventional Yukawa force shown in Equation (\ref{ar-Yukawa}) and of the power law force in Equation (\ref{ar-power}) after extending these to composition-dependent treatment.
In the case $A_{ij}\neq 1$, we introduce new parameters $\tilde{\alpha}$  \cite{dent2008eotvos} and $\tilde{n}$, as distinguished from the $\alpha$ and $n$ used in the composition-independent case $A_{ij}= 1$ \cite{CQG-Review}, as 
\begin{equation}
a(r)=\tilde{\alpha} \left(1+\frac{r}{\lambda}\right) e^{-r/\lambda},
\label{ar-Yukawa-cd}
\end{equation}
for the Yukawa parameterization, and
\begin{equation}
a(r)=(1+\tilde{n})\left( \frac{\lambda}{r} \right)^{\tilde{n}},
\label{ar-power-cd}
\end{equation}
for the power law parameterization.
Using these parameterizations, a least square analysis of the data shown in Figure \ref{result} was performed to obtain the constraints on $\tilde{\alpha}$ and $\tilde{n}$.
In this analysis, numerial integration over the material volume was performed, supposing the distance dependence of the model parametrization.

The new ``gravitational charge" $q$ defined in Equation (\ref{mod-G}) can be expressed in terms of the baryon number, e.g., as $B=Z+N$ ($Z$ and $N$ are the atomic and neutron numbers, respectively), or as $I_Z=N-Z$, and so on. 
In the case of $q = B$, we obtain
\begin{equation}
|\tilde{\alpha}_{q=B}|<5.5 \times 10^{-2}
\end{equation}
at $\lambda$ = 1 cm. This baryon-number coupling force was first proposed by Lee and Yang \cite{lee1955conservation}. 

\begin{figure}
\begin{center}
\includegraphics[clip,width=100mm]{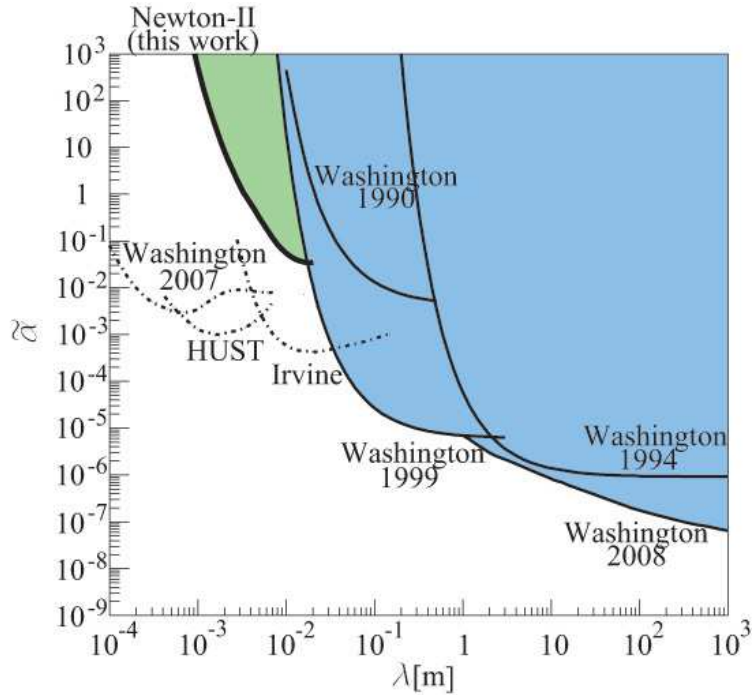}
\caption{\label{xiL} Constraints on the Yukawa coupling parameter $\tilde{\alpha}$ ($95\:  \%\: C.L.$) for various direct measurements \cite{schlamminger2008test} in the case of $q=B$, where shaded area indicates excluded area. The present study is shown as Newton-II \cite{ninomiya2013short}. Results from inverse square law tests, which are interpreted as "indirect", without testing WEP, are also plotted as dashed lines \cite{CQG-Review} (HUST \cite{PhysRevLett.108.081101}, Irvine \cite{PhysRevD.32.3084}). References for Washington are same as for Fig.\ref{eta-r}, except for Washington 1990 \cite{PhysRevD.42.3267} and 2007 \cite{PhysRevLett.98.021101}.   }
\end{center}
\end{figure}

\begin{figure}
\begin{center}
\includegraphics[clip,width=100mm]{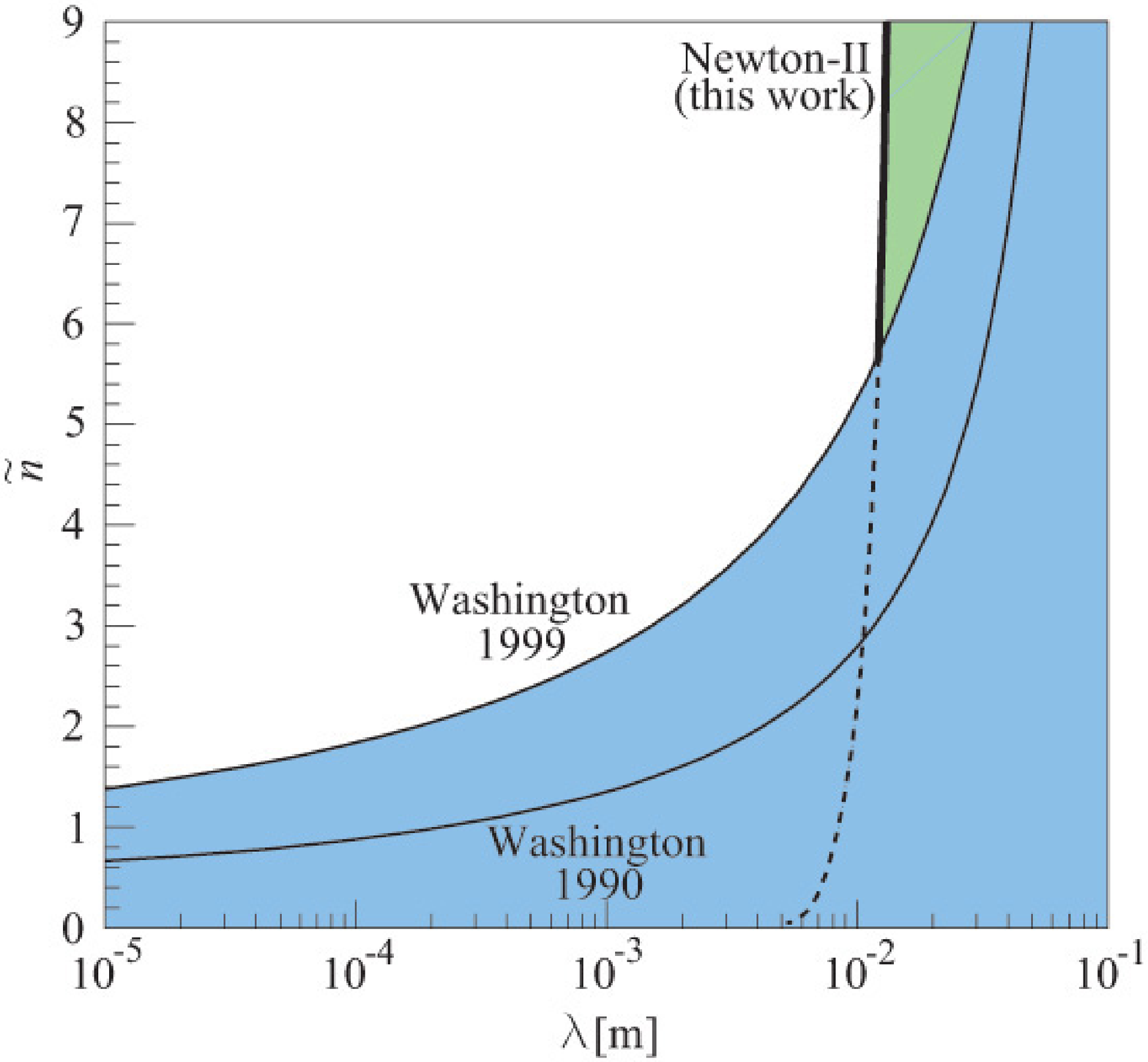}
\caption{\label{nL} Constraints on $\tilde{n}$ for various direct measurements \cite{schlamminger2008test} in the case of $q=B$, where shaded area indicates excluded area. The result of this study is shown as Newton-II. These results are obtained from each WEP violation parameter $\eta_{ij}$. References are same as for Fig.\ref{xiL}. Results from inverse square law tests cannot be shown without ambiguity because of no $\eta_{ij}$ data.}
\end{center}
\end{figure}

Experimental constraints on $\tilde{\alpha}$ and $\tilde{n}$ as a function of the range parameter $\lambda$ are shown in Figures \ref{xiL} and \ref{nL}, respectively.
Figure \ref{xiL} corresponds to the conventional $\alpha-\lambda$ plot for Yukawa parameterization for testing the gravitational inverse square law, as an extension for composition dependence.
The characteristics of this $\tilde{\alpha}-\lambda$ plot can be simply understood from the following discussion.
If we measure a composition dependence of the gravitational constant at a distance $r$ , a typical experimental quantity to be measured is the ratio
\begin{equation}
\gamma_{ij}(r)=\frac{\tilde{G}_{ik}(r)}{\tilde{G}_{jk}(r)},
\label{gamma}
\end{equation}
between objects $i$ and $k$ and objects $j$ and $k$. 
Then, constraints on possible model parameters can be obtained by solving 
\begin{equation}
\gamma_{ij}(r)=\frac{1+A_{ki} a(r)}{1+A_{kj} a(r)}.
\label{ratio}
\end{equation}
For the Yukawa parameterization of Equation (\ref{ar-Yukawa-cd}),
\begin{equation}
\tilde{\alpha}=\frac{\gamma_{ij}-1}{A_{ki}-\gamma_{ij}A_{kj}}\frac{1}{1+r/\lambda}e^{r/\lambda}
\label{model-Yukawa}
\end{equation}
gives us the constraint curve of $\tilde{\alpha}$ using the experimental value of $\gamma_{ij}$, including its measuring error.

By the definition of $\gamma_{ij}$ in Equation (\ref{gamma}), it can be shown that $\gamma_{ij}$ can be extracted from the WEP violation parameter $\eta_{ij}$ as
\begin{equation}
\gamma_{ij}=-\frac{\eta_{ij}+2}{\eta_{ij}-2},
\end{equation}
which yields $\gamma_{ij}$ directly from Equation (\ref{eta-G}).
In our present analysis, we use all the data in Figure \ref{result}, including distance dependence; therefore, the obtained precision for $\tilde{\alpha}$ is better than in this simple calculation.
Indeed, if we do not use our distance dependence data, the obtained precision decreases as
\begin{equation}
|\tilde{\alpha}_{q=B}|<3.2 \times 10^{2} \; ({\rm no \; }r{\rm\!-\!dependence}).
\end{equation}
This results from the factor $(\gamma_{Al-Cu}-1)/(A_{W-Al}-\gamma_{Al-Cu} A_{W-Cu})$ being large for our material combination.
It will be possible to improve this constraint in the future by using a materials combination with large $\Delta(B/\mu)_{ij}$, such as Be-Ti.

In addition to the Yukawa parameterization, we analyzed the results using the power law parameterization of Equation (\ref{ar-power-cd}).
The constraints on the $\tilde{n}-\lambda$ parameter space are shown in Figure \ref{nL}.
In this case, simple calculation using $\gamma_{ij}(r)$ is obtained from Equation (\ref{ratio}) as
\begin{equation}
\lambda=\left(  
\frac{1}{1+n}
\frac{\gamma_{ij}-1}{A_{ki}-\gamma_{ij}A_{kj}}
\right)^{1/\tilde{n}} r.
\label{model-power}
\end{equation}

As Equation (\ref{mod-G}) is a two-dimensional function of $A_{ij}$ and $r$, $\tilde{\alpha}$ can be examined not only by testing the composition dependence but also by measuring the distance dependence. Although $\tilde{\alpha}$ represents composition dependence, the experimental precision is dominated by the measurements of distance dependence, as discussed above. In fact, $\tilde{\alpha}$ can be constrained much tighter than in the present study by testing the inverse square law without testing the composition dependence at all \cite{PhysRevLett.98.021101,PhysRevLett.98.131104}. 
The reason that data containing only distance dependence can set constraints on the $\tilde{\alpha} - \lambda$ and $\tilde{n} - \lambda$ parameter space in Figures \ref{xiL} and \ref{nL} can be understood as follows.
By their definitions, the relationships among $\tilde{\alpha}$ and $\alpha$, and $\tilde{n}$ and $n$ are
\begin{equation}
\alpha=A_{ij} \tilde{\alpha}; \; \lambda^n=A_{ij} \lambda^{\tilde{n}}. 
\end{equation}
For the actual value of $A_{ij}$ of nearly $1$, constraint curves on $\alpha - \lambda$ and $n-\lambda$ can appear at nearly the same positions in the $\tilde{\alpha}-\lambda$ and $\tilde{n}-\lambda$ plots.
The corresponding constraints are plotted in Figures \ref{xiL} and \ref{nL}.
Inverse square law tests can set constraints not only for $\alpha - \lambda$ but also for $\tilde{\alpha}-\lambda$.
For example, if we obtain the following experimental data;
\begin{equation}
\gamma=\frac{\tilde{G_{ij}}(r_1)}{\tilde{G_{ij}}(r_2)},
\end{equation}
where $\gamma$ is the ratio of the gravitational constant measured at different distances $r_1$ and $r_2$ with a common combination of compositions $i,j$, then, 
\begin{equation}
\gamma=\frac{1+A_{ij}\tilde{\alpha}(1+r_1/\lambda)e^{-r_1/\lambda}}{1+A_{ij}\tilde{\alpha}(1+r_2/\lambda)e^{-r_2/\lambda}},
\end{equation}
which yields
\begin{equation}
\tilde{\alpha}=\frac{1}{A_{ij}}
\frac{\gamma-1}{(1+\frac{r_1}{\lambda})e^{-r_1/\lambda} - \gamma
(1+\frac{r_2}{\lambda})e^{-r_2/\lambda}
}.
\end{equation}
This is the reason why test results of the inverse square law can contribute to constrain the composition dependent parameter $\tilde{\alpha}$.
However, these ``indirect" constraints cannot inversely be interpreted as a WEP test.
In other words, $\tilde{\alpha}-\lambda$ and $\eta$ are not equivalent; and
$\eta$ cannot be obtained from the $\tilde{\alpha}-\lambda$ constraint.
In this sense, the WEP violation parameter $\eta$ should be regarded as the quantity directly representing the composition dependence of $G$.
It is interesting to point out that, tests of the inverse square law, such as \cite{PhysRevLett.98.021101}, used different material attractors to cancel Newtonian gravity.
However, such measurements did not test the composition dependence of $G$ at the same distance, therefore, $\eta$ cannot be obtained without supposing model parametrization.

Our results setting the constraints at the shortest range of around 1 cm are obtained from the direct determination of the gravitational constant for different materials, and the WEP violation parameter $\eta$.
In terms of the power law parameterization, our results set a new constraint on $\lambda$ in the large $\tilde{n} \geq 6$ region.

As a plan, not only we can still improve the experimental sensitivity by changing the test materials, but also extend our WEP study towards shorter range at around 1 mm region, by utilizing our newer apparatus Newton-IVh \cite{CQG-Review}.

\section{CONCLUSION \label{conclusion}}
In this study, we performed a direct measurement of the composition dependence of the gravitational constant $G$ at the shortest range of around 1 cm with a precision of $10^{-2}$.
The obtained results are consistent with the universality of $G$.
This result can also be interpreted as a short range test of WEP by assuming WEP at a long range.

\section{Acknowledgements \label{ack}}
This study is supported by a Grant-in-Aid for Exploratory Research (grant numbers 18654048 and 20654024), and Rikkyo SFR (Rikkyo University Special Fund for Research), MEXT-Supported Program for the Strategic Research Foundation at Private Universities, 2014-2017 (S1411024).
Parts of this study were performed as undergraduate student experiments. The authors thank Y. Miyano, M. Takahashi, T. Tsuneno, T. Amanuma, S. Danbara, T. Iino, S. Mizuno, Y. Araki, T. Ohmori, Y. Sakurai, S. Yamaoka and Y. Sekiguchi for their important inputs in this study.

\section*{References}
\bibliography{gravity}

\end{document}